\documentclass[twocolumn]{aastex631}
\usepackage{newtxtext,newtxmath}
\usepackage[T1]{fontenc}

\usepackage{graphicx,graphics}	% Including figure files
\usepackage{color,epsfig}
\usepackage{amsmath}	% Advanced maths commands
\usepackage{amssymb}	% Extra maths symbols
\usepackage{psfrag}
\usepackage[T1]{fontenc}
\usepackage{ebgaramond}
\usepackage{newtxmath}

\newcommand{\Hi}{H\,\textsc{i}}
\newcommand{\HI}{H\,\textsc{i}~}
\def\HII{H\,\textsc{i}~}

\def\bxHi{\bar{x}_{\rm H\,\textsc{i}}}

\def\k{{\bm{k}}}
\def\a{{\bm{a}}}
\def\x{{\bm{x}}}
\def\r{{\bm{r}}}
\def\U{{\bm{U}}}

\def\cl{{\mathcal C}_{\ell}}

\def\n{\hat{\bm{n}}}
\def\v{\bm{v}}

\def\l{\mathcal L}

\def\dnu{\Delta \nu}
\def\nubar{\Bar{\nu}}
\def\yhi{y_{\rm H\,\textsc{i}}}
\begin{document}
% \label{firstpage}
% \pagerange{\pageref{firstpage}--\pageref{lastpage}}
% \maketitle

\title{The Evolving Power Spectrum for the light cone Epoch of Reionization (EoR) 21-cm signal}

\correspondingauthor{Suman Pramanick}
\email{suman21eor@gmail.com}

\author[0000-0002-3665-292X]{Suman Pramanick}
% \email{sukhdeepsingh5ab@gmail.com}
\affiliation{Department of Physics, Indian Institute of Technology Kharagpur, Kharagpur 721 302, India}

\author[0000-0002-2350-3669]{Somnath Bharadwaj}
\affiliation{Department of Physics, Indian Institute of Technology Kharagpur, Kharagpur 721 302, India}

\author[0000-0003-1206-8689]{Khandakar Md Asif Elahi}
\affiliation{Centre for Strings, Gravitation and Cosmology, Department of Physics, Indian Institute of Technology Madras, Chennai 600036, India}

\author[0000-0001-7728-3756]{Rajesh Mondal}
\affiliation{Department of Physics, National Institute of Technology Calicut, Calicut 673601, Kerala, India}

\begin{abstract}
The rapid evolution of the cosmological neutral hydrogen (\Hi) distribution during the EoR is imprinted along the line of sight (LoS) in the redshifted 21-cm signal due to the light-cone (LC) effect. The LC EoR 21-cm signal ceases to be ergodic along the LoS, and the Fourier transform-based three-dimensional power spectrum (PS) fails to capture the full two-point statistics. Several earlier studies have used the multi-frequency angular power spectrum (MAPS) $\cl(\nu_1,\nu_2)$ to overcome this limitation. However, we do not have a simple interpretation of $\cl(\nu_1,\nu_2)$ in terms of comoving length scale, and the data volume is large. Here we introduce the evolving power spectrum (ePS) to quantify the two-point statistics of the LC EoR 21-cm signal. This has a simple interpretation in terms of redshift evolution and comoving length scales, and the binned ePS reduces the data volume by several orders of magnitude compared to MAPS. Considering simulations, we study the first three even angular multipoles of ePS to quantify the LoS anisotropy of the signal. We find that as reionization progresses, at large $k$ ($ \ge 0.6 \, {\rm Mpc}^{-1}$), $P_{e\,0}(k,z)$ the monopole moment decreases as $\propto\bxHi$ the mean neutral \HI fraction, which, in principle, can be used to observationally determine the reionization history. Furthermore,  $P_{e\,2}(k,z)$ the quadrupole moment is negative at small $k$ and positive at large $k$. We propose the binned ePS, which captures the entire information contained in MAPS, to quantify the full two-point statistics of the LC EoR 21-cm signal.

\end{abstract}


% %%%%%%%%%%%%%%%%%%%%%%%%%%%%%%%%%%%%%%%%%%%%%%%%%%

% %%%%%%%%%%%%%%%%% BODY OF PAPER %%%%%%%%%%%%%%%%%%

% % , SKA-LOW\footnote{\url{http://www.skatelescope.org}} \citep{Koopmans_2015} 

\section{Introduction}
\label{sec:intro}
The Epoch of Reionization (EoR) is a pivotal phase in the evolution history of the Universe. During this era, all baryonic matter transitioned from a cold, neutral state to a hot, ionized state. Analyses of quasar absorption spectra reveal a highly ionized intergalactic medium (IGM) at redshifts ($z \sim 5.5$) with a mean neutral hydrogen fraction of $\bxHi \sim 10^{-5} - 10^{-4}$ (e.g., \citealt{becker2001evidence, fan2002evolution}). These observations, along with the evolution of Ly-$\alpha$ and Ly-$\beta$ effective optical depth \citep{mcgreer2014model}, suggest a rapid rise in ionization fraction at $z > 5.5$. Furthermore, studies of the clustering and luminosity function of Ly$\alpha$ emitters (LAEs) (e.g., \citealt{ouchi2010statistics, faisst2014spectroscopic, konno2014accelerated,ota2017new}) support this rapid change. Cosmic Microwave Background (CMB) observations constrain the reionization epoch through the Thomson scattering optical depth ($\tau_{\rm Th}$). The latest measurements by the Planck telescope suggest reionization began around $z \sim 12$. Combining these diverse observations, the EoR is constrained to the redshift range $6 \lesssim z \lesssim 12$ (e.g., \citealt{robertson2013new, robertson2015cosmic, mondal2017, mitra2017cosmic,mitra2018first, dai2019constraining}), with a possibility of extending as late as $z \sim 5.5$. However, the exact timing and duration of this crucial epoch remain uncertain. The recent detection of high redshift galaxies by the James Webb Space Telescope (JWST) \citep{Gardner_2023} sheds light on the possible sources and extent of EoR \citep{ Adams2022, Castellano_2022, Austin_2023, Harikane_2023, Finkelstein_2023}. 

The most promising probe of the EoR is the 21-cm emission line from neutral hydrogen (\Hi) in its ground state. Significant efforts are currently underway to detect the redshifted 21-cm signal from the EoR. Some experiments aim to detect this in the frequency spectrum of the sky-averaged (global) signal  \citep{bowman2008toward, bowman2018absorption, singh2018saras}. Other experiments aim to detect this through the statistics of the brightness temperature fluctuations measured as a function of angle and frequency \citep{mertens2020improved,trott2020deep,pal2020demonstating,Abdurashidova2022hera,Chatterjee2024tracking}. The latter use low-frequency radio interferometers such as the Giant Metrewave Radio Telescope (GMRT)\footnote{\url{http://www.gmrt.ncra.tifr.res.in}} \citep{swarup1991, Paciga2013}, Low-Frequency Array (LOFAR)\footnote{\url{http://www.lofar.org}} \citep{van2013lofar, yatawatta2013initial}, Murchison Widefield Array (MWA)\footnote{\url{http://www.haystack.mit.edu/ast/arrays/mwa}} \citep{bowman2013science, tingay2013murchison, dillon2014overcoming}, and several others. However, the 21-cm signal is intrinsically weak, and is masked by much stronger foreground emissions that exceeds the expected signal by several orders of magnitude (e.g. \citealt{ali2008foregrounds, ghosh2012characterizing}). Despite significant advancements in foreground mitigation techniques, definitive detection remains elusive. The current best upper limits on the 21-cm power spectrum come from the Hydrogen Epoch of Reionization Array (HERA)\footnote{\url{http://reionization.org}} \citep{furlanetto2009cosmology, deboer2017hydrogen}, reaching 
$457~\text{mK}^2$ at $k=0.34~h\;\text{Mpc}^{-1}$ at $z=7.9$ and $3496~\text{mK}^2$ at $k=0.36~ h\;\text{Mpc}^{-1}$ at $z=10.4$ \citep{Abdurashidova_2023}.

In this paper, we focus on one aspect of the EoR 21-cm signal, namely the light cone (LC) effect \citep{barkana2006light,zawada2014light}. Our view of the redshifted 21-cm signal is restricted to the backward light cone, wherein the look-back time increases with distance along the line-of-sight (LoS). The statistical properties of the \HI distribution evolve rapidly as reionization proceeds, and the EoR 21-cm signal ceases to be statistically homogeneous  along the LoS. As a consequence, the three-dimensional (3D) power spectrum (PS) fails to fully quantify the statistics of the observed EoR 21-cm signal  \citep{datta2012light, datta2014light,trott2016exploring}.  \citet{mondal2018} have proposed the  multi-frequency angular power spectrum (MAPS), $\cl(\nu_1,\nu_2)$ \citep{datta2007multifrequency} to accurately quantify the two-point statistic of LC EoR 21-cm signal. Several subsequent studies have used MAPS  to  analyse the statistics of simulated LC EoR 21-cm signal.  \citet{mondal2019method} have proposed a method to observationally determine the reionization history of our Universe using MAPS, \citet{mondal2020} have made predictions for measuring  MAPS using the upcoming SKA-Low radio interferometer and \citet{mondal2022} consider the prospects of constraining reionization parameters using MAPS. 

Despite its ability to completely quantify the two-point statistics of the LC EoR 21-cm signal, MAPS  $\cl(\nu_1,\nu_2)$ suffers from two major drawbacks. First, the data volume is very large, this being of the order of the number of frequency channels squared for each angular multipole $\ell$. Second, it is difficult to directly interpret $\cl(\nu_1,\nu_2)$ in terms of comoving length-scales that pertain to physical features in the \HI distribution. In this paper we introduce the evolving power spectrum (ePS) that is related to MAPS through a change of variables, followed by a Fourier transform.  We show that the binned ePS is able to quantify the entire information that is contained in MAPS. It has the advantage that the data volume is much smaller in comparison to MAPS. Further, it is straightforward to interpret the ePS in terms of redshift evolution and comoving length-scales.  

Redshift space distortion (RSD) due to peculiar velocities is an important effect that is imprinted in the 21-cm signal \citep{bharadwaj2004cosmic}. This causes the EoR 21-cm PS to be anisotropic with respect to the LoS direction, and several works quantify this using simulation \citep{mao2012redshift, jensen2013probing, majumdar2013effect}. However, these are restricted to snapshots that have a fixed redshift, they do not incorporate the LC effect. In this paper, we show that it is possible to quantify RSD in LC simulations using the angular multipole moments of the ePS. We also show that, at small scales, the redshift evolution of the monopole moment of the ePS, traces the evolution of the mean neutral hydrogen fraction.  

This paper is structured as follows -- In section~\ref{sec:statistics}, we discuss the statistics of the EoR 21-cm signal. We discuss the mathematical formulation of the evolving power spectrum in subsection~\ref{sec:eps}. In section~\ref{sec:sim}, we discuss the EoR 21-cm LC simulations used here. Section~\ref{sec:results} discusses the main results of this paper, and finally, we summarise and discuss our findings in section~\ref{sec:conc}.

This paper has used Plank+WP best-fitting values of the cosmological parameters $h=0.6704$, $\Omega_{\rm m0} = 0.3183$, $\Omega_{\Lambda 0} = 0.6817$, $\Omega_{\rm b 0} h^2 = 0.022 032$, $\sigma_8 = 0.8347$ and $n_{\rm s} = 0.9619$ \citep{collaboration2020planck}.

\section{Statistics of EoR 21-cm signal}
\label{sec:statistics}
The quantity of interest here is the 21-cm brightness temperature fluctuations $\delta T_{\rm b}(\hat{\bf n}, \nu)$, where $\hat{\bf n}$ refers to a direction on the sky and $\nu$ is the observed frequency of the redshifted 21-cm radiation. The frequency $\nu$ not only encodes the comoving distance $r$ where the radiation originated, but also the radial component of the peculiar velocity at the position ${\bf r}=r \hat{\bf n}$ \citep{bharadwaj2004cosmic}. Furthermore, our view of the Universe is restricted to the backward light cone $r=c (\eta_0-\eta)$, where, in terms of conformal time, $\eta$ is when the 21-cm signal originated and $\eta_0$ is the present value. This implies that each frequency $\nu$ corresponds to a different look-back conformal time $\eta$ \citep{barkana2006light}. The LC effect is particularly important during the EoR when the mean \HI fraction $\bar{x}_{\HI}$ and the \HII bubble size distribution both evolve rapidly along the radial direction. This causes the statistical properties of $\delta T_{\rm b}(\hat{\bf n}, \nu)$ to vary rapidly along $\nu$ \citep{mondal2018}. 

Although the EoR 21-cm signal ceases to be statistically homogeneous (or ergodic) along $\nu$ due to the LC effect, it is still statistically homogeneous and isotropic on the celestial sphere. We quantify the two-point statistics of $\delta T_{\rm b}(\hat{\bf n}, \nu)$ using $\cl(\nu_1,\nu_2)$ the multi-frequency angular power spectrum (MAPS) defined using   \citep{datta2007multifrequency} 
\begin{equation}
    \delta T_{\rm b} ({\bf \hat{n}}, \nu) = \sum_{\ell, m} a_{\ell m }(\nu) Y^m_\ell ({\bf \hat{n}}) \,,
\end{equation}
and 
\begin{equation}
    \cl(\nu_1,\nu_2) = \left<a_{\ell m} (\nu_1) a^*_{\ell m} (\nu_2)\right> \,.
\end{equation} 

In the present paper, we restrict our analysis to a small region of the sky that subtends a small solid angle $\Omega \ll 1$. We may then work in the "flat sky approximation" where the sky is a two-dimensional (2D) plane, and we have a fixed line of sight (LoS) or radial direction, which is orthogonal to the plane.   Here we express the 21-cm signal as  $\delta T_{\rm b}({\mathbf \theta}, \nu)$  where ${\mathbf \theta}$ is a two-dimensional vector on the plane of the sky. We expand  $\delta T_{\rm b}({\mathbf \theta}, \nu)$ in Fourier modes  
\begin{equation}
    \Delta \tilde{T} ({\bf \ell}, \nu) = \int{d^2 \theta} e^{2\pi i {\bf \ell} \cdot   {\mathbf \theta}} \delta T_{\rm b}({\bf \theta}, \nu).
\end{equation}
and define the MAPS as 
\begin{equation}
    \cl (\nu_1,\nu_2) = \Omega^{-1} \left< \Delta \tilde{T}({\bf \ell},\nu_1) \Delta \tilde{T}^{*}(\mathbf{\ell},\nu_2) \right>.
\end{equation}

\subsection{The statistics of a snapshot 21-cm signal}
\label{sec:snapshot_stat}
A "snapshot" refers to the redshifted 21-cm brightness temperature signal $\delta T_{\rm b} ({\bf r},z)$ from a three-dimensional volume at a fixed instant of look-back time, or equivalently redshift $z$. For brevity of notation, in this subsection, we do not explicitly show $z$ as this has a fixed value for a snapshot signal. Although in reality, it is not possible to observe the snapshot signal, it serves as a good approximation when the radial extent of the volume is sufficiently small so that we may ignore the evolution of the 21-cm signal during the light travel time. In this case, we decompose $\delta T_{\rm b} ({\bf r})$ into Fourier modes using Equation (7) of \cite{bharadwaj2005using}
\begin{equation}\label{eq:tb}
\delta T_{\rm b} ({\bf r})=\int \frac{d^3 k}{(2 \pi )^3}e^{-i {\bf k}.r \hat{\bf n}} \Delta \Tilde{T}_b ({\bf k})   \,, 
\end{equation}
and we define the three-dimensional (3D) power spectrum (PS) $P({\bf k})$ through 
\begin{equation}\label{eq:psz}
    \langle \Delta \Tilde{T}_b ({\bf k}) \Delta \Tilde{T}_b^{*} ({\bf k}') \rangle = (2 \pi)^3 \delta_{\rm D}^3 ({\bf k}-{\bf k}') P({\bf k})\,,
\end{equation}
where $\delta_{\rm D}^3$ is the 3D Dirac delta function.

Note that this assumes the 21-cm signal to be statistically homogeneous or ergodic along all three spatial directions. Further, the assumption of isotropy implies $P({\bf k})=P(k)$ {\it i.e.} the PS does not depend on the direction of ${\bf k}$. However, redshift space distortion (RSD) of the 21-cm signal \citep{bharadwaj2004cosmic} introduces anisotropy along the line of sight (LoS) direction ${\bf \hat{n}}$, and the PS  $P({\bf k})=P(k,\mu) \equiv P(k_\perp, k_\parallel)$ where $k_\perp$ and $k_\parallel$ are, respectively, the components of ${\bf k}$  perpendicular and parallel to ${\bf \hat{n}}$ and $\mu=k_\parallel/k$. 
We quantify this anisotropy using \citep{hamilton1992measuring} 
\begin{equation}\label{eq:multipole_ps}
    P(k,\mu) = \sum_{q \; \rm even} {\mathcal L}_q (\mu) P_q(k),
\end{equation}
where ${\mathcal L}_q (\mu)$ are the Legendre polynomials and $P_q(k)$ are different angular multipoles of $P(k,\mu)$. In the plane parallel approximation, the effect of RSD is invariant under ${\bf \hat{n}} \rightarrow -{\bf \hat{n}}$, and consequently the odd multipoles are predicted to be zero \citep{hamilton1992measuring, hamilton1998assl, Shaun1994, Bharadwaj_1999}. Guided by this, we have considered only the even multipole moments in Equation~(\ref{eq:multipole_ps}). 

Considering the MAPS $ \cl(\nu_1,\nu_2)$ for a snapshot signal, statistical homogeneity along the LoS direction implies that this only depends on the frequency separation $\Delta \nu=\mid \nu_1-\nu_2 \mid$, and we have \citep{datta2007multifrequency} 
\begin{equation}\label{eq:cldeltanu}
\cl(\Delta \nu) = r^{-2} \, \int \frac{d k_\parallel}{(2 \pi)}  e^{-i  k_\parallel r' \Delta \nu}  P(k_\perp, k_\parallel) \, 
\end{equation}
where $r$ is the comoving distance corresponding to $z$, $r^{'}=dr/d \nu$ and $k_\perp= \ell /r$. In summary, the statistic of the snapshot 21-cm signal is equally well described by the PS $P(k_\perp, k_\parallel)$ or the MAPS $\cl(\Delta \nu)$.

\subsection{The evolving power spectrum}
\label{sec:eps}
The 3D PS  $P(k)$ (Equation~(\ref{eq:psz})) is very useful to quantify and interpret the statistics of the EoR 21-cm signal. For example, we can interpret $\Delta^2(k)=k^3 P(k)/2 \pi^2$ as the mean-squared brightness temperature fluctuations on the comoving length scale $2 \pi/k$. However, this assumes the signal to be statistically homogeneous along the LoS. \citet{mondal2018} show that $P(k)$ is inadequate to fully quantify the two-point statistics of the EoR 21-cm signal due to the LC effect. It is necessary to consider the full MAPS $\cl(\nu_1,\nu_2)$  which does not assume the signal to be statistically homogeneous along the LoS. Unfortunately, we do not have a simple physical interpretation for $\cl(\nu_1,\nu_2)$ in terms of the comoving length scales. Furthermore,  it is quite voluminous (square of the number of frequency channels for each $\ell$), although the signal is mainly localized within a narrow band around the diagonal $(\nu_1 \approx \nu_2)$. In this work, we introduce the evolving power spectrum (ePS) $P_e(k,z)$ (or $P_e(k,\mu,z) \equiv P_e(k_\perp, k_\parallel,z)$) that captures the entire information in the MAPS $\cl(\nu_1,\nu_2)$. The ePS has the advantage that we can interpret it in terms of comoving length scales, just like the PS $P(k)$. The ePS is also considerably less voluminous compared to the MAPS. 

We proceed by performing a change of variables. We express $\cl(\nu_1,\nu_2)$ as $\cl(\Delta \nu,\bar{\nu})$, where 
\begin{equation}\label{eq:rescaling}
\Bar{\nu} = \frac{\nu_1+\nu_2}{2} \;{\rm and} \; \dnu = \nu_1 - \nu_2 .
\end{equation}
Here, $\cl(\Delta \nu,\bar{\nu})$ would be independent of $\bar{\nu}$ if the EoR 21-cm signal were statistically homogeneous along the LoS. The $\bar{\nu}$ dependence captures the fact that the statistical properties of the EoR 21-cm signal evolve along the LoS.  We can relate the mean frequency $\bar{\nu}$ to a redshift $z$. We use this to define the ePS through  
\begin{equation}\label{eq:eps}
\cl(\Delta \nu,\bar{\nu}) = r_z^{-2} \, \int \frac{d k_\parallel}{(2 \pi)}  e^{-i  k_\parallel r_z' \Delta \nu}  P_e(k_\perp, k_\parallel,z) \,,  
\end{equation}
which is exactly the same as Equation~(\ref{eq:cldeltanu}), except for the $\bar{\nu}$ dependence in the L.H.S., and the $z$ dependence in the R.H.S.. Note that $P_e(k_\perp,k_\parallel,z)$ is not defined through Equation~(\ref{eq:psz}) that assumes the signal to be statistically homogeneous in all three spatial directions. The ePS $P_e(k_\perp,k_\parallel,z)$, here, is defined through Equation~(\ref{eq:eps}), and it captures the entire information that is contained in $\cl(\nu_1,\nu_2)$. Furthermore, we use Equation~(\ref{eq:multipole}) to evaluate the multipole expansion of $P_e(k_\perp,k_\parallel,z)$, and consider various multipole moments, namely the monopole $P_e(k,z) \equiv P_{e\,0}(k,z)$, quadrupole $ P_{e\,2}(k,z)$ and hexadecapole  $P_{e\,4}(k,z)$.

\section{Simulating the LC EoR 21-cm Signal}
\label{sec:sim}

Our LC simulations involve two main steps. In the first step, we simulate the snapshot hydrogen ionization field for several closely spaced redshifts. We start by using an $N$-body code to simulate the matter and halo distributions at eight redshifts in the range $z=7.56$ to $z=8.71$. We have interpolated the gridded matter and halo density fields at a finer redshift interval and used the semi-numerical code {\sf ReionYuga}\footnote{\url{ https://github.com/rajeshmondal18/ReionYuga}} \citep{mondal_2025_14948653} to generate $57$ snapshots of the ionization field across the redshift range mentioned above. 
In the second step, we extract slices from these snapshots, stitch them together, and then apply RSD to construct the LC EoR 21-cm signal.  We refer the reader to \cite{pramanick2023quantifying} for details of the simulation methodology. 

Our simulations have a comoving volume of $[286.72 \, {\rm Mpc}]^3$ and a minimum halo mass of $1.09 \times 10^9 \, M_{\sun}$, which are limited by the 1 TB RAM available  on our computer. We note that this value of the minimum halo mass for the ionizing sources is consistent with the findings of \cite{Finlator2016min}.
The comoving volume, minimum halo mass, and the values of the astrophysical parameters used for the present simulations are very similar to those used in several earlier works \citep{ mondal2018, mondal2019method, mondal2021epoch,  pramanick2023quantifying}.  Considering the complete history of reionization for these simulations, earlier works show that reionization starts at $z \sim 13$ and ends at $z \sim 6$, with $50\%$ ionization at $z \approx 8$.  These simulations yield an integrated Thomson scattering optical depth $\tau = 0.057$, consistent with the Planck Collaboration's observed value of $\tau = 0.058 \pm 0.012$ \citep{collaboration2020planck}.

Our final LC EoR 21-cm simulation corresponds to a redshift (frequency) range $z_n =7.53 \; (\nu_n = 165.92 \;{\rm MHz})$ to $z_f = 8.51 \; (\nu_f = 149.26 \;{\rm MHz})$, centered at $z_c = 8.01 \; (\nu_c = 157.64 \;{\rm MHz})$.  Here the subscripts $n$ and $f$ refer to the near and far ends along the LoS of the box respectively. The frequency bandwidth $B=17.06 \;{\rm MHz}$ is divided into $N_c = 512$ channels of resolution $\dnu = 0.033 \; {\rm MHz}$. The comoving span of the LC box along the LoS is from $r_n = 9013.42$ to $r_f = 9300.14$.  The mean neutral hydrogen fraction varies from $[\bxHi]_f=0.65$ to $[\bxHi]_n=0.34$ along the LoS of the LC simulation. We note that the comoving distance $r$ only changes by $3 \%$ across the LoS extent of the simulation box, in contrast to $\bxHi$ which changes by $50 \%$. Here, we have ignored the change in $r$ and used fixed values $r = 9149.86\, {\rm Mpc}$ and $r' = dr/d\nu = 16.79  \;{\rm Mpc \, MHz^{-1}}$  that correspond to the central redshift $ z_c$.  

\begin{figure}
\centering
\includegraphics[width=0.48\textwidth]{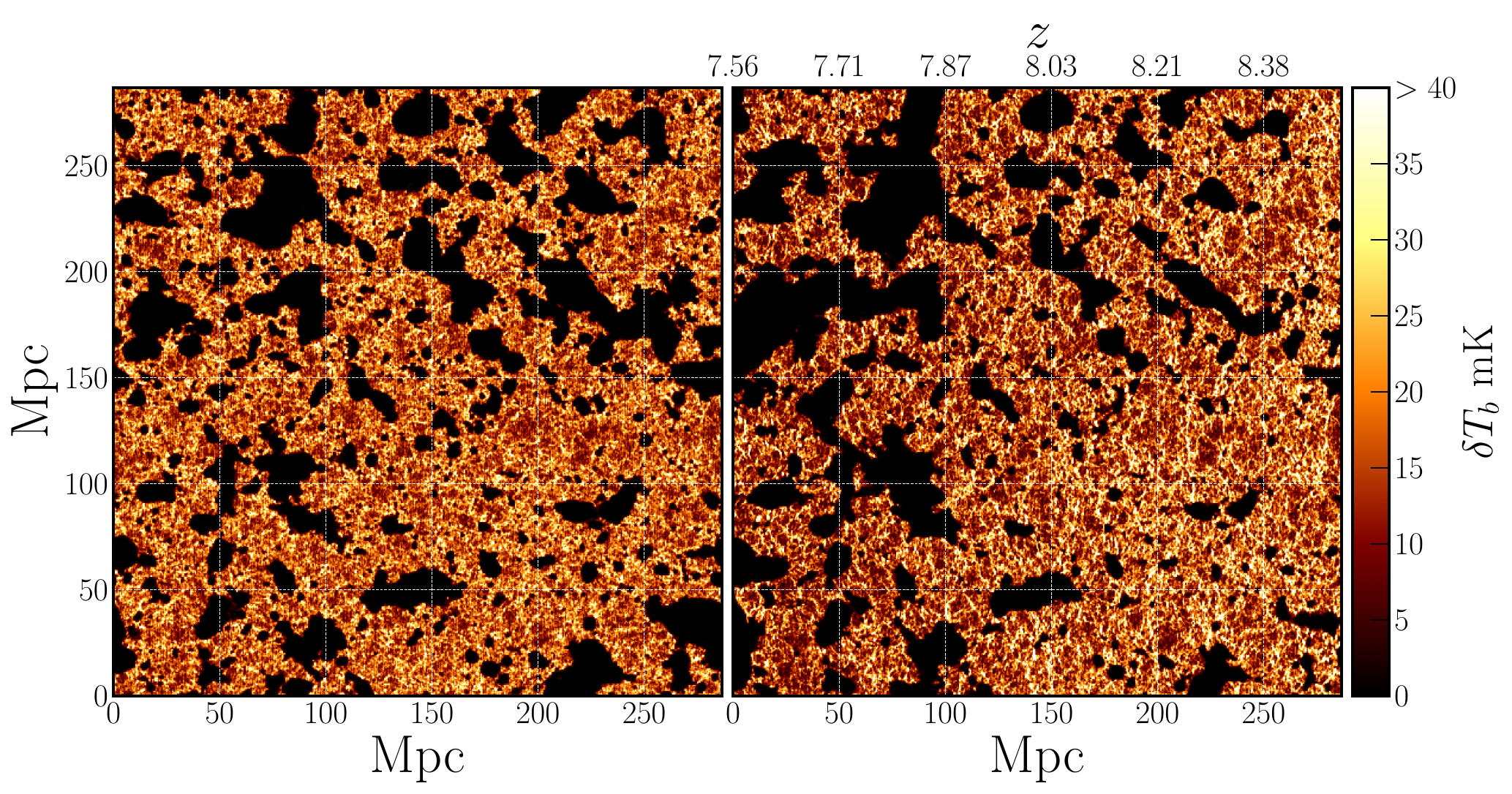}
\caption{The 21-cm brightness temperature maps, $\delta T_{\rm b}$, for a snapshot simulation (left panel) and an LC simulation (right panel), both have a central redshift of $z_c=8$.  Note the redshifts $z$ shown along the upper $x$ axis, which is the LoS direction in the LC simulation.}
\label{Tb}
\end{figure}

We have generated $18$ statistically independent realizations of the LC simulations. Each realization has a different seed for the Gaussian random field (initial density fluctuations) used in the $N$-body simulation. We used these realizations to calculate the mean and $1 \sigma$ errors for the results shown in the subsequent section. Figure \ref{Tb} shows a comparison between a snapshot (left) and an LC (right) simulation. In contrast to the snapshot, the LC simulation shows a systematic evolution of ionized bubble sizes along the LoS direction ($x$-axis). Note that the typical bubble size in the LC simulation gets larger from right to left ({\it i.e.} as reionization proceeds).

\section{Results}
\label{sec:results}
\begin{figure}
\centering
\includegraphics[width=0.48\textwidth]{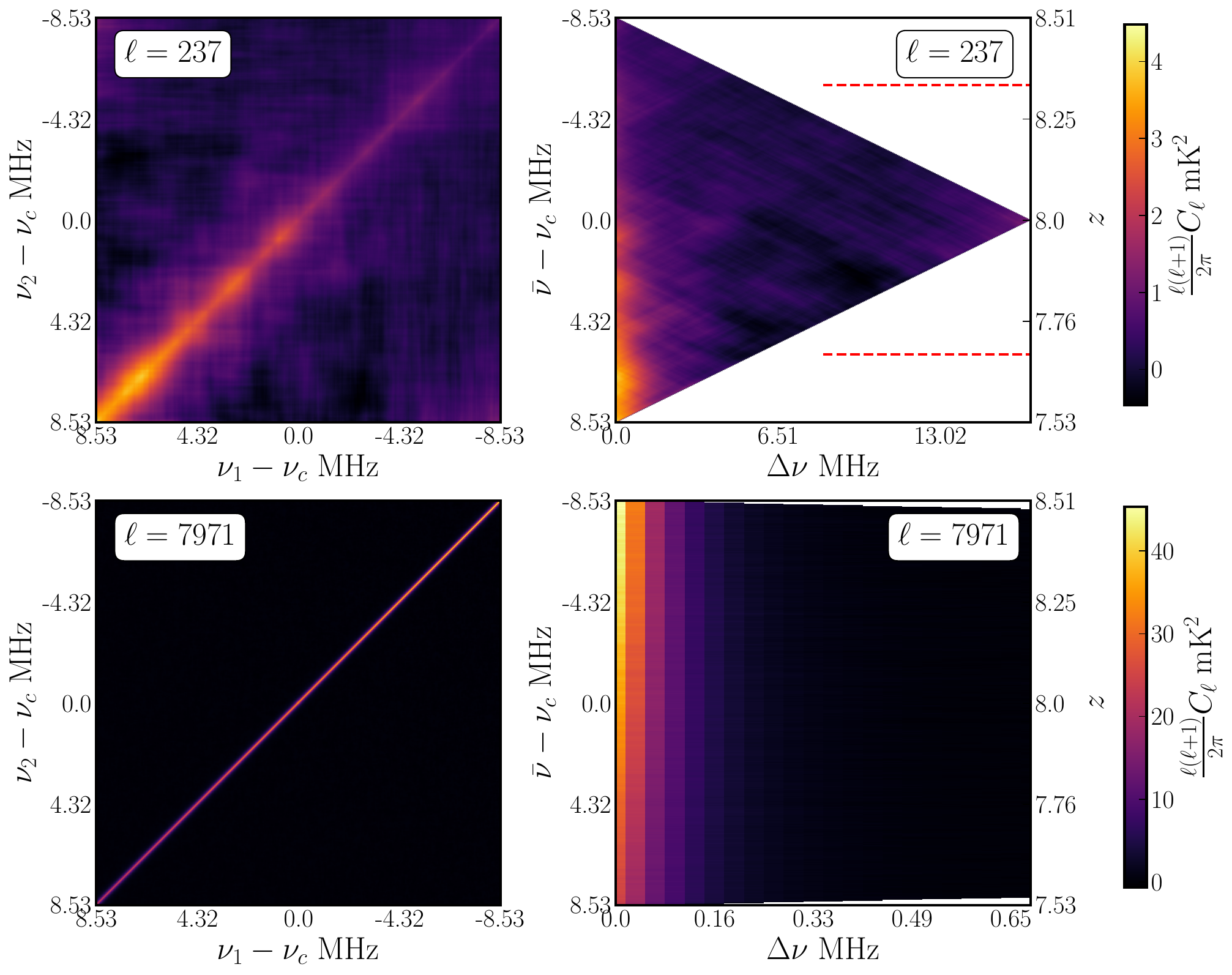}
\caption{This shows the  MAPS, the left column  shows $\cl(\nu_1, \nu_2)$ whereas the right column shows $\cl(\dnu,\nubar)$. Results are shown for  $\ell=237$ and $\ell=7971$ in the top and bottom rows respectively. For larger $\ell$, the signal de-correlates rapidly with increasing $\dnu$ and is $\sim 0$ for large $\dnu$. To account for this, we have limited the  $\dnu$ range that is shown for  $\cl(\dnu,\nubar)$ at $\ell=7971$. The red dotted lines in the upper right panel demarcate  $z=7.69,  \,{\rm and}\; 8.35$ for which the results have been shown in Figure~\ref{fig:ps_monok} and Figure~\ref{fig:psmp_k}.}
\label{fig:cl}
\end{figure}

In this section, we show the results of the LC simulation. We first discuss the top row of Figure~\ref{fig:cl}. The left panel shows the MAPS $\cl (\nu_1,\nu_2)$ for one of the smaller angular multipoles $\ell=237$. One can see the signal is maximum around the main diagonal $\nu_1 = \nu_2$ and falls rapidly away from the diagonal. In other words, the information is mainly localized near the diagonal, and there is very little information at large frequency separation which has values close to zero. The right panel shows $\cl(\dnu, \nubar)$, calculated using Equation~(\ref{eq:rescaling}) for the same $\cl (\nu_1,\nu_2)$ as shown in the left panel. The x-axis shows the values of $\dnu$,  which range from $0 - 17.06 \;{\rm MHz}$. The y-axis shows the values of $\nubar - \nu_c$ where $\nubar$ ranges from $\nu_n$ to $\nu_f$. The corresponding redshift values are shown on the right-hand y-axis. The triangular shape of the plot can be understood as follows. The bottom left corner corresponds to $\nu_1 = \nu_2 = \nubar = \nu_n$, for which we have only a single value $\Delta \nu=0$. The number of available $\Delta \nu$ values increases with $\nubar$, and it is the highest ($N_c/2$) for $\nubar=\nu_c$. The number of available $\Delta \nu$ values decreases beyond $\nubar=\nu_c$, and again we have a single value $ \Delta \nu=0$ in the upper left corner that corresponds to $\nu_1=\nu_2=\nubar=\nu_f$.  It is clearly visible that for all $\nubar$, the values of $\cl(\dnu, \nubar)$  peak at $\dnu=0$ and fall rapidly with increasing $\dnu$. We see that the amplitude of the signal increases as reionization proceeds (top to bottom along the y-axis in the figure). 

The bottom row of Figure~\ref{fig:cl} shows the same quantities as the top row, but for a larger angular multipole $\ell = 7971$. In the left panel, we see that the values of $\cl (\nu_1,\nu_2)$ are very sharply localized near the main diagonal, and the values are close to zero for $\Delta \nu \ge 0.15 \, {\rm MHz}$. This motivates us to show $\cl(\dnu, \nubar)$ for only a small range of  $\Delta \nu$ values in the right panel. In this case, we see that the amplitude of the signal decreases as reionization proceeds. Comparing the two rows of the figure, we notice three interesting features in $\cl(\dnu, \nubar)$. First, for both angular multipoles, $\cl(\dnu,\nubar)$ falls rapidly with increasing $\dnu$. Further, this de-correlation occurs faster for larger $\ell$ values. As noted in several earlier works \citep{bharadwaj2001hi, bharadwaj2005using}, this is a generic feature of the cosmological 21-cm signal through all phases of its evolution. Second, for all $\ell$, we find that $\cl(\dnu,\nubar)$ varies slowly with $\nubar$ compared to its rapid de-correlation with respect to $\dnu$. This allows us to reduce the data volume by smoothing $\cl(\dnu,\nubar)$ with respect to $\nubar$ without loss of information. Here, for the subsequent analysis, we have collapsed $16$ consecutive $\nubar$ channels. However,  we maintain the original channel resolution along $\dnu$ to capture the rapid $\dnu$ variation. Third, we find that the amplitude of $\cl(\dnu,\nubar)$ changes considerably with respect to $\nubar$ (or $z$). Note that, we expect $\cl(\dnu,\nubar)$ to be independent of $\nubar$ for an ergodic signal. The $\nubar$ dependence seen here clearly illustrates that the EoR 21-cm signal ceases to be statistically homogeneous (or ergodic) along the LoS due to the LC effect. Interestingly, the nature of the $\nubar$ dependence changes with $\ell$. We see that for small $\ell$ (top row) the amplitude of $\cl(\dnu,\nubar)$ increases as reionization progresses, whereas the reverse is observed for large $\ell$ (bottom row). The transition between these two different $\nubar$ dependence seems to occur around $\ell \sim  2600 $ (not shown here). Possible causes for this feature are discussed later after we consider the ePS which shows a similar behaviour.

 % Naively, we expect the amplitude of $\cl(\dnu,\nubar)$  to be determined by $\bxHi$, which decreases as reionization proceeds. However,  the fluctuations in the \HI distribution, which are length-scale dependent,   increase as reionization proceeds. The evolution  of $\cl(\dnu,\nubar)$  is decided by a combination of these two competing 

\begin{figure}
\centering
\includegraphics[width=0.48\textwidth]{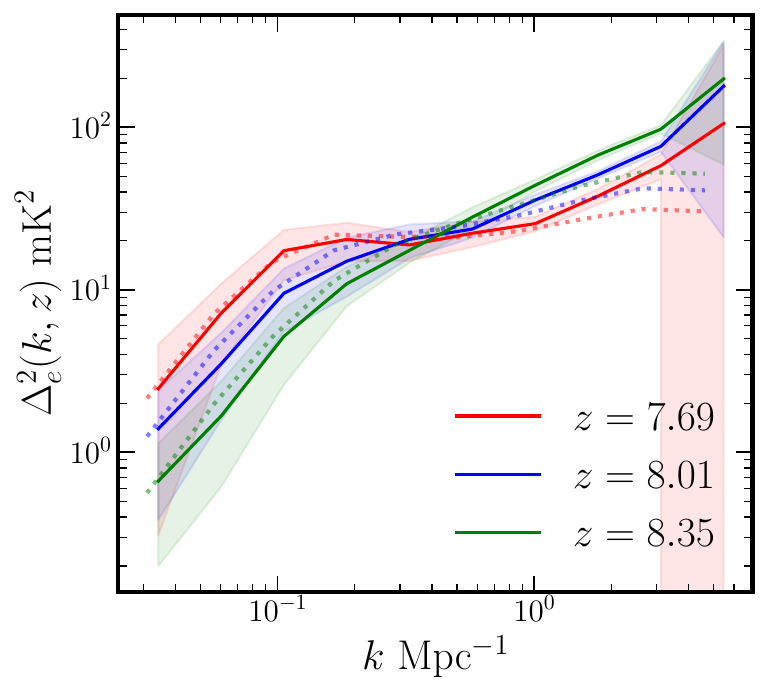}
\caption{This shows the mean squared brightness temperature fluctuations $\Delta^2_e (k, z)$ for three different redshifts $z=7.69, \, 8.01 \,{\rm and}\; 8.35$. Solid lines show $\Delta^2_e (k, z)$, estimated from ePS $P_e(k,z)$, whereas the dotted lines show the same estimated from EoR 21-cm snapshot PS. Shaded regions show $1 \sigma$ error bars for the ePS. The largest $k$ bin, $k=5.47\,{\rm Mpc}^{-1}$ is noisy and show large fluctuations.}
\label{fig:ps_monok}
\end{figure}

\subsection{Monopole only}
We now consider the ePS that we have estimated from the $\cl(\dnu, \nubar)$ shown in Figure \ref{fig:cl}. We first consider $P_e(k,z)$ the spherically binned ePS where we ignore the LoS anisotropy of the 21-cm signal, and express the cylindrical ePS as 
\begin{equation}\label{eq:pscy-mono}
P_e(k_\perp,k_\parallel,z)=P_e(k,z) \,.
\end{equation}
In the subsequent discussion, 
we refer to this case as "Monopole only". The $(k_\perp,k_\parallel)$ plane was divided into $10$ spherical bins of equal logarithmic spacing in $k$. Considering a particular bin with mean comoving wave number $k_a$,   $P_e(k_\perp,k_\parallel,z)$ has the same value $P_e(k_a,z)$ for all the  $(k_\perp,k_\parallel)$  modes within this bin. The spherically binned ePS $P_e(k,z)$ for the $10$ bins are each related to $\cl(\Delta \nu,\bar{\nu})$ through Equation~(\ref{eq:eps}) and Equation~(\ref{eq:pscy-mono}). Following \citet{elahi2023a}, for each central reshift $z$, we have used least-squares fitting to directly estimate $P_e(k,z)$ from the $\cl(\Delta \nu,\bar{\nu})$ determined from the LC simulations.\footnote{We have used eqs. (15) and (16) of \citet{elahi2023a}, assuming the noise covariance to be an identity matrix.}

Figure~\ref{fig:ps_monok} shows the mean squared brightness temperature fluctuations $\Delta^2_e(k, z) = k^3 P_e(k, z)/2\pi^2 $ for three different redshifts, $z=7.69, \, 8.01 \,{\rm and}\; 8.35$. The solid lines show $\Delta^2_e(k, z)$ estimated from $P_e(k, z)$ the ePS, whereas the dotted lines show the same estimated from the snapshot PS, while the shaded regions show the respective $1 \sigma$ error bars (for ePS). The three redshifts have been chosen so that $z=8.01=z_c$ corresponds to the central redshift of our LC simulation, which is also the middle stage of reionization where $\bxHi=0.50$. The other two redshifts, $z=8.35$ and $7.69$, which are demarcated on the right $y$ axis of the upper right panel of Figure~\ref{fig:cl}, respectively correspond to $\bxHi=0.60$ and $\bxHi=0.39$. Considering the ePS, we see that at all $z$,  $\Delta^2_e(k, z)$ increases with $k$  at small $k$, flattens out at $k \sim 0.3-0.6 \, {\rm Mpc}^{-1}$ and then again increases  with $k$ at large $k$. Considering the $z$ dependence, $\Delta^2_e(k, z)$ decreases with increasing $z$  at small $k$,  whereas the reverse is true at large $k$. The transition occurs at $k \sim 0.3-0.6 \, {\rm Mpc}^{-1}$ where all the $\Delta^2_e(k, z)$ intersect.  Considering the snapshot PS, the results are very similar to those for the ePS, except at large $k$ where  $\Delta^2_e(k, z)$ increases rapidly with $k$ for the ePS in contrast to the snapshot PS where $\Delta^2(k, z)$ increases gradually with $k$. The ePS has much more power at small length-scales $(k >0.6 \, {\rm Mpc}^{-1})$ in comparison to the snapshot PS. Comparing the snapshot and LC simulations shown in Figure~\ref{Tb}, the gradual evolution of the neutral fraction possibly manifests itself as an enhancement in the small-scale fluctuations in the \HI distribution.

\begin{figure}
\centering
\includegraphics[width=0.48\textwidth]{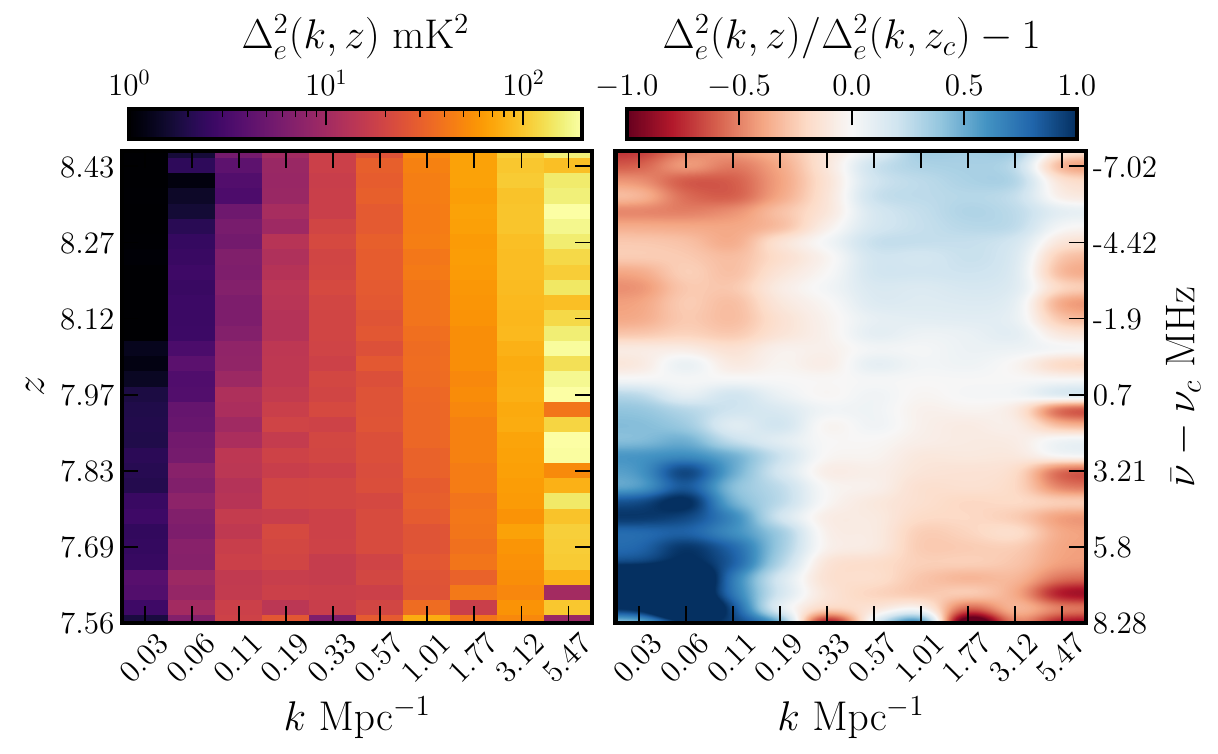}
\caption{Left panel  shows the mean squared brightness temperature fluctuations $\Delta^2_e (k, z)$ for the entire $z$ and $k$ range of our simulation. The right panel shows  $\Delta^2_e (k, z)/\Delta^2_e(k, z_c)-1$, for which we have applied Gaussian smoothing in $k$ and $z$ to mitigate the rapid fluctuations and highlight the smoothly varying component.}
\label{fig:ps_mono_heat}
\end{figure}

\begin{figure*}
\centering
\includegraphics[width=\textwidth]{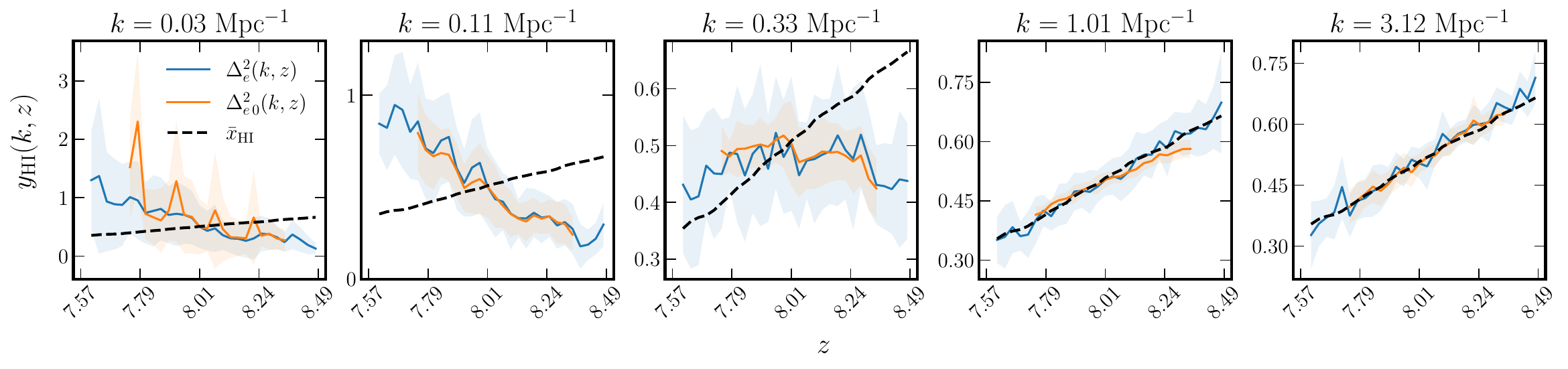}
\caption{The blue and orange solid lines show $\yhi(k,z)$ for $\Delta_e^2(k,z)$ and $\Delta_{e\,0}^2(k,z)$ respectively, the shaded regions show the corresponding $1\sigma$ error bars and the black dashed line shows the evolution of the mean neutral hydrogen fraction $\bxHi(z)$ (reionization history). Each panel corresponds to a different value of $k$ indicated on the top of the panel.}
\label{fig:pk_z}
\end{figure*}

The left panel of Figure~\ref{fig:ps_mono_heat} shows $\Delta^2_e(k, z)$ (from ePS) for the entire $z$ range of our simulation. We see that for all $z$,  $\Delta^2_e(k, z)$ increases with increasing $k$. For small $k$ ($ < 0.3 \, {\rm Mpc}^{-1}$), it is clearly visible that $\Delta^2_e(k, z)$ increases as reionization progresses (top to bottom). However, at large $k$ ($k \ge 0.6 \, {\rm Mpc}^{-1})$, the redshift evolution of $\Delta^2_e(k, z)$ is not very clearly visible from this figure. To clearly visualize the redshift evolution of $\Delta^2_e(k, z)$, we consider $[\Delta^2_e(k, z)/\Delta^2_e(k, z_c)]-1$  (right panel of Figure~\ref{fig:ps_mono_heat}) which shows the change relative to the value at $z_c$. We see that for small $k$ ($ < 0.3 \, {\rm Mpc}^{-1}$) the value of $\Delta^2_e(k, z)$ increases by $\sim 67  \%$ from $z=8.4$ to $z_c$, and increases further by  $\sim 150 \%$ from  $z_c$ to $z=7.6$. At intermediate $k$ ($0.3 - 0.6 \, {\rm Mpc}^{-1}$) the value of  $\Delta^2_e(k, z)$ is roughly independent of $z$. At large $k$  ($ > 0.6 \, {\rm Mpc}^{-1}$) the value of $\Delta^2_e(k, z)$ decreases by $\sim 30\%$ from $z=8.4$ to $z_c$, and decreases further by $\sim 36\%$ from  $z_c$ to $z=7.6$. Note that in the subsequent discussion, we denote the three distinct $k$ ranges (length-scales) referred to above as the small (large), intermediate (intermediate), and large (small) $k$ values (length-scales), respectively. Note further that the largest k value is at the tip of the diagonal in the 3-dimensional k space. As a consequence, the largest k bin does not contain many $(k_\perp, k_\parallel)$ modes, which is why the estimated ePS is noisy in this bin. We thus exclude $k=5.47\;{\rm Mpc}^{-1}$ bin from our analysis.
% Note further that we exclude the largest $k$-bin, which is found to be noisy. 

To further analyze the redshift evolution of ePS, we define a quantity 
\begin{equation}\label{eq:yhi}
\yhi(k,z) = \bxHi(z_c) \frac{\Delta^2_e(k,z)}{\Delta^2_e(k,z_c)}
\end{equation}
shown in Figure~\ref{fig:pk_z} for five different $k$ values. Note that $\yhi(k,z)$ has been defined so that $ \yhi(k,z_c)=\bxHi(z_c)$ i.e. it matches the mean neutral fraction at $z_c$. The blue solid lines show $\yhi(k,z)$, the blue shaded regions show the corresponding $1 \sigma$ error bars, and the black dashed lines show $\bxHi (z)$.  The first two panels show the evolution of $\yhi(k,z)$ at large length scales. We see that $\yhi(k,z)$ decreases with $z$, the opposite evolution compared to  $\bxHi(z)$. The middle panel shows $\yhi(k,z)$ for the intermediate length scale.  Here we see that $\yhi(k,z)$ is roughly flat, independent of $z$.  The last two panels show $\yhi(k,z)$ for small length scales. Here we find that $\yhi(k,z)$ increases with $z$, and it matches the evolution of $\bxHi (z)$ (i.e. $\yhi(k,z)=\bxHi(z)$) across the $z$ range considered here. 
% We note that this behavior is similar to the findings of \citet{mondal2019method}, and 
This holds the potential to trace the evolution of $\bxHi(z)$ from the measurements of the ePS.

\begin{figure*}
\centering
\includegraphics[width=0.92\textwidth]{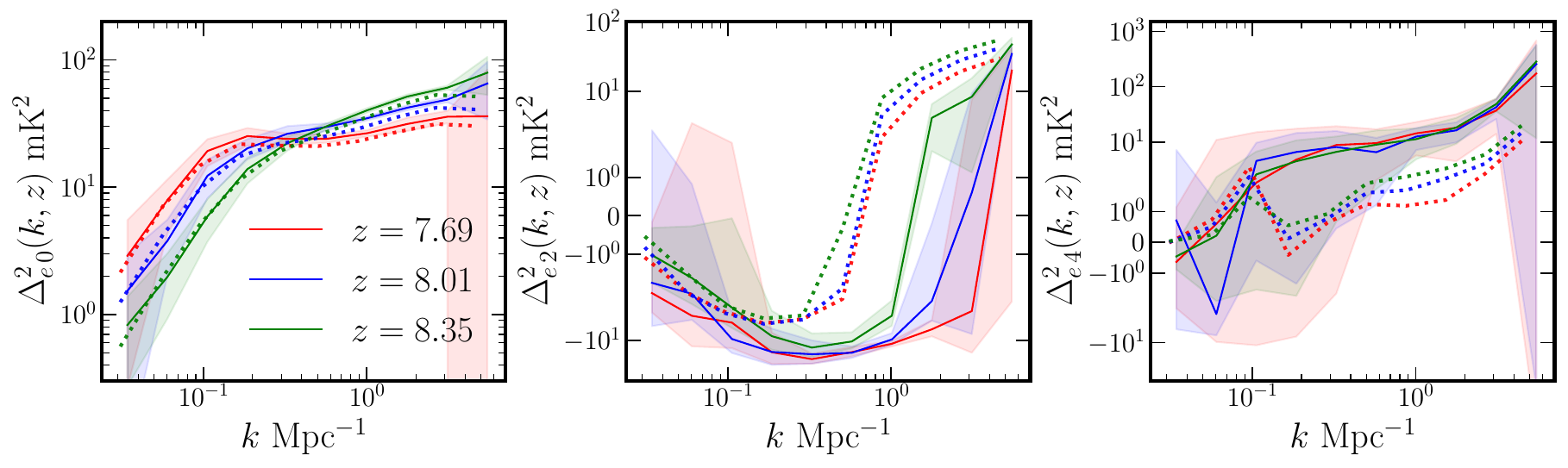}
\caption{This shows the ePS multipole moments $\Delta_{e \, q}^2(k, z)$ as a function of $k$ for the monopole ($q=0$, left panel), quadrupole ($q=2$, middle panel) and hexadecapole ($q=4$, right panel). The results are shown for three $z$ values as indicated. Solid lines show $\Delta_{e\, q}^2(k, z)$, shaded regions show corresponding $1 \sigma$ error bars and dotted lines show the multipole moments $\Delta_{q}^2(k, z)$ for the respective snapshot PS.}
\label{fig:psmp_k}
\end{figure*}

\subsection{Higher Multipoles}
Here we consider the LoS anisotropy of the LC EoR 21-cm signal. In addition to RSD, it is also possible that the ePS may acquire some LoS anisotropy because of the LC effect itself. To quantify the anisotropy, we expand the cylindrical ePS using 
\begin{equation}\label{eq:multipole}
P_e(k_\perp, k_\parallel, z) \equiv P_e(k,\mu, z) = \sum_{q \; \rm even} {\mathcal L}_{ q} (\mu) \, \, P_{e \, q}(k,z),
\end{equation}
where $P_{e \, q}(k,z)$ are the  multipole moments of the cylindrical ePS. As mentioned earlier (Equation~\ref{eq:multipole_ps}), all the odd multipoles are predicted to be zero. In the subsequent discussion, we refer to this case as "Higher multipole". The $(k_\perp,k_\parallel)$ plane was divided into $10$ spherical bins of equal logarithmic spacing in $k$. The binned multipole moments of the ePS $P_{e \, q} (k,z)$ for these $10$ bins are related to $\cl(\Delta \nu,\bar{\nu})$ through Equation~(\ref{eq:eps}) and Equation~(\ref{eq:multipole}). 
% {\bf Here we have directly estimate $P_{e \, q}(k,z)$ by fitting $\cl(\dnu,\nubar)$ to some Fourier mode for each $z$ and $k$, where the wave number of the mode $k$ is determined by $k_\perp = \ell/r$ and associated $k_\parallel$ pair using $k=\sqrt{k_\perp^2+k_\parallel^2}$.} 
% {\bf We use similar equation as Equation~\ref{eq:eps_matrix} with the change in $[B_q]_{i}(a,n) = \sum_m {\mathcal L}_{q} (\mu) A_{nm}$. We use this to directly estimate $P_{e \, q}(k,z)$ from the $\cl(\Delta \nu,\bar{\nu})$.}
Here, we have used least-squares fitting to directly estimate $P_{e \, q}(k,z)$ from the $\cl(\Delta \nu,\bar{\nu})$. 
In this work, we only consider the first three even multipoles of the ePS, namely the monopole $P_{e \, 0}(k,z)$, the quadrupole $P_{e \, 2}(k,z)$, and the hexadecapole $P_{e \, 4}(k,z)$. The higher multipoles, expected to be much smaller, have been ignored here.

The different panels of Figure~\ref{fig:psmp_k} shows the multipole moments $\Delta_{e \, q}^2(k, z) = k^3 P_{e\, q}(k, z)/2\pi^2$ as a function of $k$ for the same three redshifts as in Figure~\ref{fig:ps_monok}. In all panels, the solid lines show the results for the ePS from the LC simulation, the dashed lines show the same multipole moments for the PS from the snapshot simulations, and the shaded regions show the corresponding $1\sigma$ error bars for the ePS. We first discuss the left panel that considers $q=0$ the monopole. We see that at small and intermediate $k$ $(< 0.6 \, {\rm Mpc}^{-1})$, the results for $\Delta_{e\,0}^2(k, z)$ are very close to those for $\Delta_{e}^2(k, z)$ (Figure~\ref{fig:ps_monok}) which has already been discussed earlier. Considering $k> 0.6 \, {\rm Mpc}^{-1}$, we have seen that $\Delta_{e}^2(k, z)$ increases rapidly with $k$. Here we find that $\Delta_{e\,0}^2(k, z)$ shows a similar behavior, but increases less rapidly with $k$. Comparing $\Delta_{e\,0}^2(k, z)$ with $\Delta_{0}^2(k, z)$ from the snapshot simulations (dotted lines), we see that the two are in close agreement throughout. Note the difference from Figure~\ref{fig:ps_monok} where $\Delta_{e}^2(k, z)$  exceeds the snapshot predictions at large $k$ $(> 0.6 \, {\rm Mpc}^{-1})$.

The middle panel of Figure~\ref{fig:psmp_k} shows the results for $q=2$ the quadrupole moment. Note that a positive quadrupole moment indicates an elongation along LoS, while a negative value indicates a squashing. The reader is referred to Figure 6 of \citet{Hamilton1998} for a detailed interpretation of the quadrupole and hexadecapole moments.
We see that for all $z$, $\Delta_{e\,2}^2(k, z)$ is negative at small and intermediate 
$k$ $(< 0.6 \, {\rm Mpc}^{-1})$. At small $k$, $\Delta_{e\,2}^2(k, z)$ decreases with increasing $k$,  then saturates at intermediate $k$ $(0.3- 0.6 \, {\rm Mpc}^{-1})$, and subsequently increases with $k$ at large $k$ $(>0.6 \, {\rm Mpc}^{-1})$. We have positive values of $\Delta_{e\,2}^2(k, z)$ at the largest $k$. The exact values of $k$ corresponding to the negative to positive transition vary in the range $ 1 < k < 3 \, {\rm Mpc}^{-1}$, and decrease with increasing $z$. We also note that $\Delta_{e\,2}^2(k, z)$ increases with $z$ for most values of $k$, however, this is more pronounced at large $k$ as compared to small $k$. Considering $\Delta_{2}^2(k, z)$ for the snapshot simulation, we see that the behavior is similar to $\Delta_{e\,2}^2(k, z)$. However, the $k$ range where the negative values of $\Delta_{2}^2(k, z)$ saturate, and the $k$ value corresponding to the negative to positive transition, both shift to smaller $k$.

The right panel of Figure~\ref{fig:psmp_k} shows $\Delta_{e\,4}^2(k, z)$ the hexadecapole moment of the ePS. This has a large scatter and is consistent with zero at small and intermediate $k$ $(<0.6 \, {\rm Mpc}^{-1})$. We find a $k$ range  $(0.6-4.0 \, {\rm Mpc}^{-1})$ where  $\Delta_{e\,4}^2(k, z)$ increases with $k$ and has positive values that are well above zero. However, $\Delta_{e\,4}^2(k, z)$ does not exhibit any significant $z$ dependence. Considering  $\Delta_{4}^2(k, z)$ for the snapshot simulations, the behavior is somewhat similar, but the values are smaller and are mostly consistent with zero.

\begin{figure}
\centering
\includegraphics[width=0.48\textwidth]{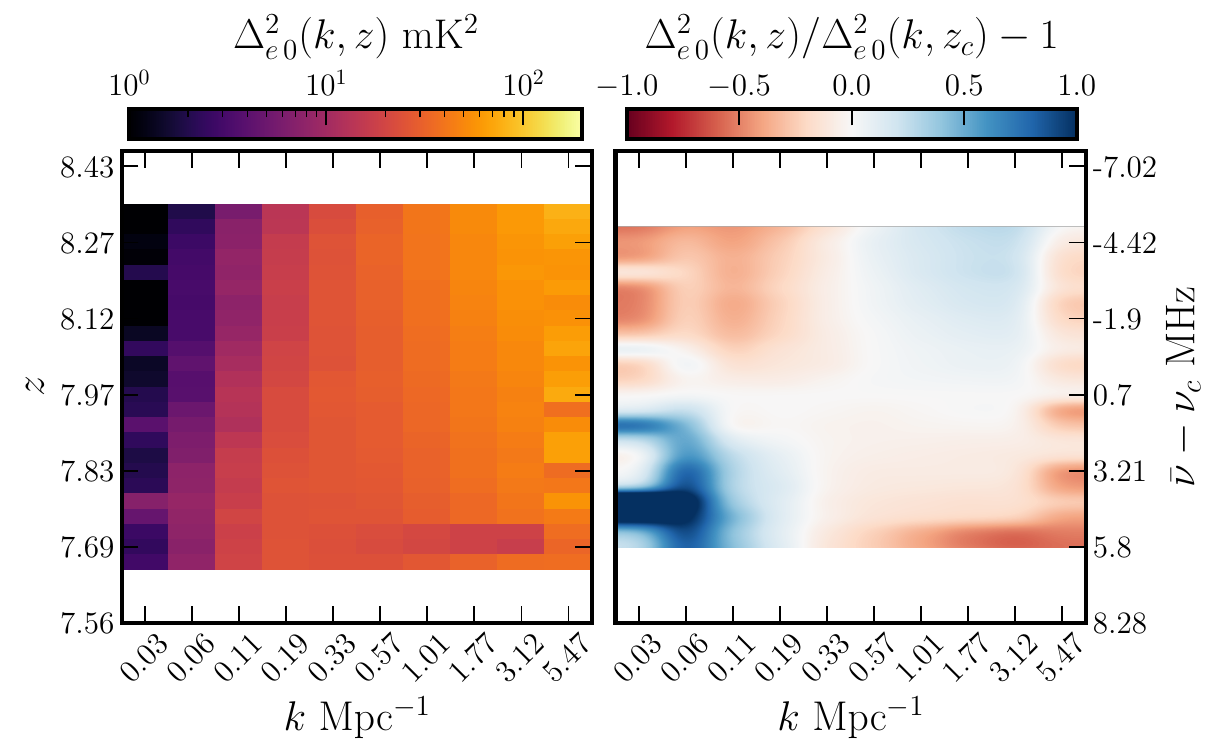}
\caption{Left panel  shows the mean squared brightness temperature fluctuations $\Delta^2_{e \, 0}(k, z)$ corresponding to the monopole moment of the ePS.  The right panel shows $\Delta^2_{e \, 0} (k, z)/\Delta^2_{e \, 0}(k, z_c)-1$, for which we have applied Gaussian smoothing in $k$ and $z$ to mitigate the rapid fluctuations and highlight the smoothly varying component.}
\label{fig:ps_mp_heat_1}
\end{figure}

Figure~\ref{fig:ps_mp_heat_1} is very similar to Figure~\ref{fig:ps_mono_heat}, except that it shows $\Delta_{e\,0}^2(k, z)$ instead of $\Delta_{e}^2(k, z)$. Note that the $\Delta \nu$ range that is available in $\cl(\dnu,\nubar)$ shrinks when $\nubar$ approaches the near and far ends of the LC simulations (Figure~\ref{fig:cl}). This makes it difficult to reliably estimate the ePS multipole moments at these two ends.  Here we have discarded $2.08 \, {\rm MHz}$ in $\nubar$ from both ends, which appears as blank regions in Figure~\ref{fig:ps_mp_heat_1}. We see that at small $k$, $\Delta_{e\,0}^2(k, z)$  increases as $z$ decreases (reionization progresses),  at intermediate $k$, $\Delta_{e\,0}^2(k, z)$ is roughly independent of $z$, whereas at large $k$, $\Delta_{e\,0}^2(k, z)$ decreases as $z$ decreases. Overall, the behaviour of $\Delta_{e\,0}^2(k, z)$ is very similar to that of $\Delta_{e}^2(k, z)$ shown in  Figure~\ref{fig:ps_mono_heat}.

\begin{figure}
\centering
\includegraphics[width=0.48\textwidth]{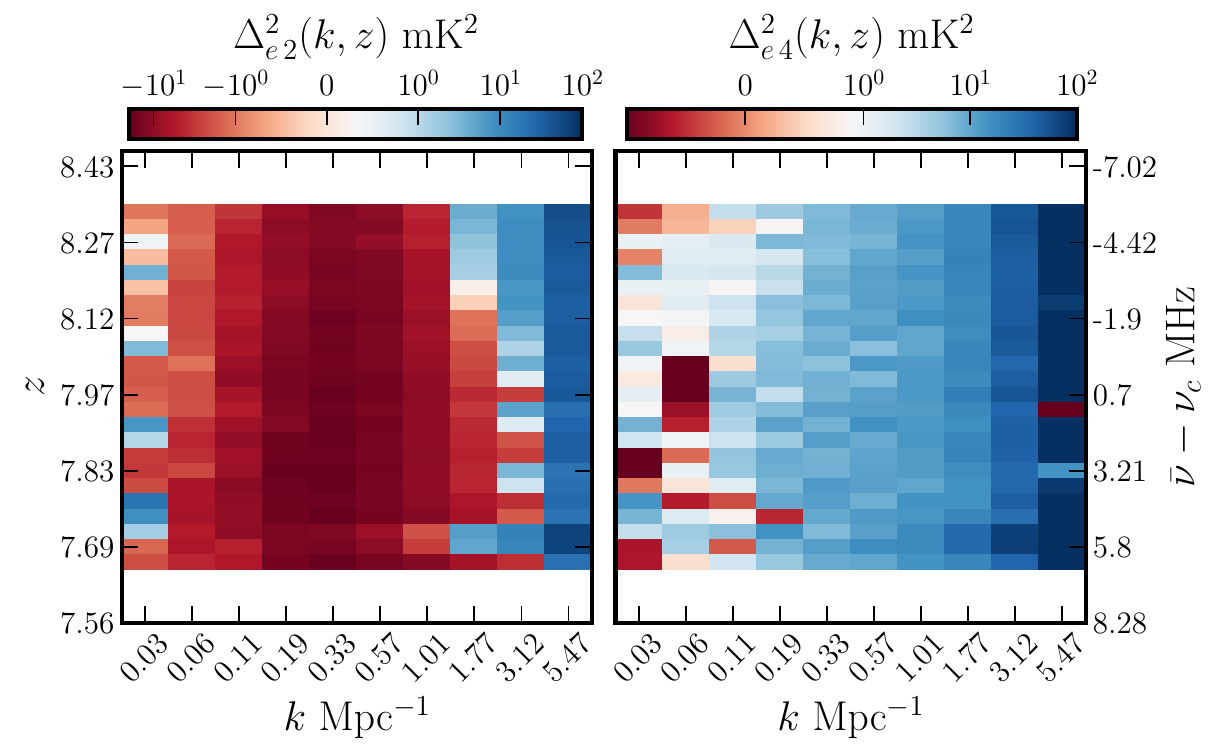}
\caption{Left panel shows $\Delta_{e\,2}^2(k, z)$ calculated for the ePS quadrupole moment $P_{e\, 2}(k,z)$ and right panel shows $\Delta_{e\,4}^2 (k, z)$ calculate for the ePS hexadecapole moment $P_{e\, 4}(k,z)$. }
\label{fig:ps_mp_heat_2}
\end{figure}

The left and right panels of Figure~\ref{fig:ps_mp_heat_2} show $\Delta_{e\,q}^2(k, z)$ as a function of $(k,z)$ for $q=2$ and $4$ respectively. We see that for all $z$, $\Delta_{e\,2}^2(k, z)$ is negative for small and intermediate $k$, and is positive at large $k$. The negative to positive transition shifts to smaller $k$ at higher $z$. The value of  $\Delta_{e\,2}^2(k, z)$ increases with $z$ at large $k$, however, it is difficult to identify a consistent $z$ dependence at smaller $k$. Considering  $\Delta_{e\,4}^2(k, z)$, the values have a large scatter at small $k$ and also at the largest $k$ bin. Considering the range  $0.3-4.0 \, {\rm Mpc}^{-1}$, we see that the values are positive and they increase with $k$. However, we do not notice any significant $z$ dependence. 

The solid orange lines in Figure~\ref{fig:pk_z} show $\yhi(k, z)$ (Equation~(\ref{eq:yhi})) calculated using $\Delta_{e\,0}^2(k, z)$ instead of $\Delta_{e}^2(k, z)$ (solid blue lines).  We see that the results from $\Delta_{e\,0}^2(k, z)$ are very similar to those from $\Delta_{e}^2(k, z)$. Both increase with $z$ at large $k$, but  $\Delta_{e}^2(k, z)$  appears to provide a better match to $\bxHi$ compared to  $\Delta_{e\,0}^2(k, z)$. However, the differences between  $\Delta_{e\,0}^2(k, z)$ and $\Delta_{e}^2(k, z)$ are within the $1\sigma$ error bars ar large $k$. 

The MAPS $\cl(\nu_1,\nu_2)$ that was originally used to quantify the statistics of the LC EoR 21-cm signal contains $N_c \times N_c=(512)^2$ data elements for each of the ten $\ell$ bins. In this paper, we have proposed the ePS, which has much less data volume, as a method to quantify the same signal. In the present analysis, we have collapsed $16$ consecutive $\nubar$ to reduce the number of channels along $\nubar$ or $z$ from $512$ to $32$. Further, as mentioned earlier, we divide the available $k$ range into $10$ bins of equal logarithmic spacing. The data volume is $32 \times 10$ when we only consider the monopole $P_e(k,z)$ (Equation~(\ref{eq:pscy-mono})) and the data volume is $24\times10\times3$ when we also include the higher multipoles of the ePS  (Equation~(\ref{eq:multipole})). Note that, for higher multipoles, we have discarded $2.08 \, {\rm MHz}$ from the two ends along the $\nubar$ (or $z$). For both cases, the data volume is much lower compared to that of the MAPS $\cl(\nu_1,\nu_2)$.  Here we assess how well the ePS, which has a much smaller data volume, is able to capture the information contained in MAPS $\cl(\dnu,\nubar)$.

\begin{figure}
\centering
\includegraphics[width=0.48\textwidth]{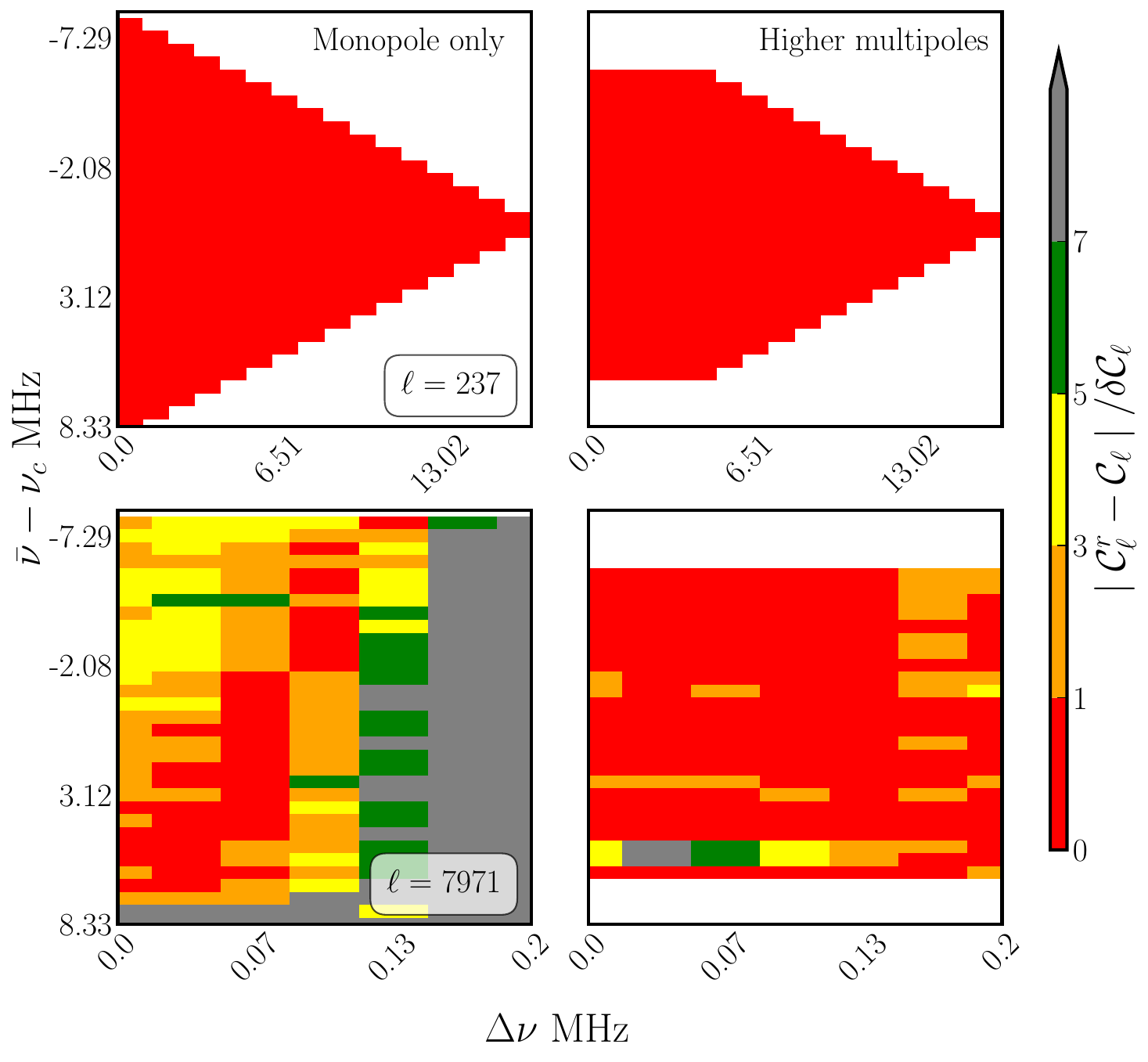}
\caption{This shows $\mid\cl^r-\cl\mid/\delta \cl$ for $\ell = 237$ in the top row and $\ell = 7971$ in the bottom row. The left column shows the monopole-only case, whereas the right column shows the higher multipole case. The top row shows the full $\dnu$ range, whereas the bottom row shows only the $\dnu$ range for which $\cl(\dnu,\nubar)$ falls $90\%$ of its maximum value (at $\dnu=0$). The difference $\mid\cl^r-\cl\mid$ is within $1\sigma$ for $\ell=237$ (top row), whereas considerable deviation at level $>5\sigma$ is observed at the bottom left panel. The difference $\mid\cl^r-\cl\mid$ is within $1-3\sigma$ in the bottom right panel.}
\label{fig:chi}
\end{figure}

We use the estimated ePS to reconstruct the  MAPS $\cl^r(\dnu, \nubar)$ by performing the Fourier transform in Equation~(\ref{eq:eps}). Figure~\ref{fig:chi} shows $\mid\cl^r - \cl \mid/\delta \cl$, which is the ratio of  $\mid\cl^r - \cl\mid$ the difference between the reconstructed MAPS and that estimated directly from the LC simulations and $\delta \cl$ the $1 \sigma$ error bars for the MAPS estimated from the LC simulations. The top and bottom rows show the results for $\ell = 237$ and $7971$, respectively, while the left column considers monopole only, whereas the right also includes higher multipoles. We see that for both panels in the top row, the deviations between $\cl^r$ and $\cl$ are within the $1 \sigma$ error bars. However, we find considerable deviations at a level $> 5 \sigma$ at $\ell = 7971$ when we consider only the monopole. The deviations are considerably reduced and are largely within $ 3 \sigma$ when we also include the higher multipoles. Note that for large $\ell$,  $\cl(\dnu, \nubar)$ falls rapidly with increasing $\dnu$, and the lower panels only show the $\dnu$ range where  $\cl(\dnu, \nubar)$   falls to $0.1$ times its maximum value. The results here indicate that the monopole ePS alone is adequate to fully capture the information in the MAPS at low $\ell$, whereas it is necessary to include the higher multipoles at larger $\ell$. This behavior is also borne out at the other $\ell$ for which the results have not been shown here. Overall, we find that the ePS is quite adequate to capture the entire information contained in the MAPS.

\section{Summary and Discussion} 
\label{sec:conc}
The statistics of the observed EoR 21-cm signal are expected to evolve significantly along the LoS direction due to the LC effect. As a consequence, the 3D PS, which assumes the signal to be statistically homogeneous along the LoS direction, fails to fully quantify the two point statistics.  
In this work,  we introduce the evolving power spectrum (ePS) to quantify the two-point statistics of the LC EoR 21-cm signal. This does not assume the signal to be statistically homogeneous along the LoS direction. Using simulations, we show that the binned ePS is capable of capturing the entire signal present in MAPS $\cl(\nu_1,\nu_2)$. The ePS has the advantage that the data volume is significantly smaller compared to MAPS, and it can be easily interpreted in terms of redshift evolution and comoving length scales. Further, we consider the higher angular multipoles of ePS to quantify the LoS anisotropy caused by RSD. 

Our LC EoR 21-cm simulations, of comoving volume $(286.7 \, {\rm Mpc})^3$, span the redshift range $z_f = 8.51$ to $z_n =7.53$   where the mean neutral \HI fraction evolves from $[\bxHi]_f=0.65$ to $[\bxHi]_n=0.34$. We have used $18$ realizations of the simulations to estimate the mean and $1 \sigma$ error bars presented here. In our analysis, we have considered two separate cases for the ePS, namely "Monopole only" and "Higher multipoles".

In Monopole only, we ignore the LoS anisotropy and only consider $P_e(k,z)$ to model the cylindrical ePS (Equation~(\ref{eq:pscy-mono})). Considering the $k$ dependence (Figure~\ref{fig:ps_monok}), we find that at all $z$,  $\Delta^2_e(k, z)$ increases with $k$  at small $k$, flattens at $k \sim 0.3-0.6 \, {\rm Mpc}^{-1}$ and then increases again with $k$ at large $k$ . Considering the $z$ dependence (Figure~\ref{fig:ps_mono_heat}), we find that $\Delta^2_e(k, z)$ decreases with increasing $z$  at small $k$,  whereas the reverse is true at large $k$. The transition occurs at $k \sim 0.3-0.6 \, {\rm Mpc}^{-1}$.

In Higher multipoles, we account for the LoS anisotropy and consider the first three even multipoles to model the cylindrical ePS (Equation~(\ref{eq:multipole})). Considering 
Figure~\ref{fig:psmp_k}, we see that the monopole $\Delta^2_{e\,0}(k,z)$ is very similar to $\Delta^2(k,z)$ shown in Figure~\ref{fig:ps_monok} for Monopole only. 
We find that for all $z$, the quadrupole $\Delta^2_{e\,2}(k,z)$ is negative for small $k$ and becomes positive at large $k$. We also find that $\Delta_{e\,2}^2(k, z)$ increases with $z$ for most values of $k$, however, this is more pronounced at large $k$ as compared to small $k$. 
Our results are qualitatively in agreement with \citep{majumdar2016effects} who have studied the redshift evolution of the quadrupole moment of the 3D PS  (Equation~(\ref{eq:psz})) for a variety of reionization models. Considering $\Delta^2_{e\,4}(k,z)$ the hexadecapole, we find that this is mostly positive, increases slowly with $k$ and  does not show any significant evolution with $z$. 

To test whether ePS is capable of fully capturing the information contained in MAPS $\cl(\dnu,\nubar)$, we reconstructed MAPS $\cl^r(\dnu,\nubar)$ from the estimated binned ePS and compared this with $\cl(\dnu,\nubar)$ (Figure \ref{fig:chi}). We find that  Monopole only and Higher multipoles are both capable of fully capturing the information contained in MAPS at small $\ell$. However, Monopole only fails to some extent at large $\ell$ whereas Higher multipoles is able to capture the full information in MAPS even at large $\ell$.  While the MAPS and the binned ePS both fully quantify the entire two-point statistics of the LC EoR 21-cm signal,  note 
the data volume, which is $(512)^2 \times 10$ for MAPS, is now reduced by a factor of $\approx 3640$  to $24\times10\times3$ for the ePS with Higher multipoles.

As noted earlier, the statistical properties of the LC EoR 21-cm signal change along the LoS direction due to the rapid redshift evolution of the mean Hydrogen neutral fraction $\bxHi$. We finally discuss the possibility of determining the redshift evolution of $\bxHi$ from the statistics of the LC EoR 21-cm signal. We note an earlier work \citep{mondal2019method,mondal2019correction} that proposes to do this using the amplitude of $\cl(\nu_1,\nu_2)$ with $\nu_1=\nu_2$.  The analysis of the present paper shows that at large $k$, $\Delta^2_e(k, z)$ the monopole moment of the ePS 
increases with $z$. Here we propose that the ePS monopole moment can be a potential tool to estimate the reionization history of the Universe. To demonstrate this, we suitably scale $\Delta^2_e(k, z)$ to define a quantity $\yhi (k,z)$ (Equation~(\ref{eq:yhi})) whose redshift evolution (at large $k$) closely matches $\bxHi$ (Figure~\ref{fig:pk_z}). Naively, one may expect the amplitude of $\Delta^2_e(k, z)$ to scale as $\bxHi^2$  in some $k$ range. However, we do not see this behaviour at any length-scale, and we find a $\bxHi$ dependence instead. This is possibly a competition between $\bxHi^2$  and the bubble size and number distribution, although the exact explanation is not known at present.  We plan to study this, in addition to the observational prospects in future work.

\section*{Acknowledgements}
We thank the anonymous reviewer for their valuable comments. Authors acknowledge the super-computing facilities at the Centre for Theoretical Studies, Department of Physics, IIT Kharagpur.

% %%%%%%%%%%%%%%%%%%%%%%%%%%%%%%%%%%%%%%%%%%%%%%%%%%
\section*{Data Availability}

The simulated data and package involved in this work will be shared on reasonable request to the authors.

\bibliography{main}{}
\bibliographystyle{aasjournal}

\label{lastpage}
\end{document}